\documentclass[proceedings, preprint]{rmaa}

\usepackage{paralist}
\usepackage{subfigure}

\usepackage{psfrag,color}

\SetYear{2010} \SetConfTitle{XIII Latin American Regional IAU
Meeting}

\title{Asteroseismology of the Delta Scuti star V650 Tauri}

\author{
  L. Fox Machado,\altaffilmark{1}
  R. Michel,\altaffilmark{1}
  M. \'Alvarez,\altaffilmark{1}
  J.N. Fu,\altaffilmark{2}
  and C. Zurita,\altaffilmark{3}}

\altaffiltext{1}{Observatorio Astron\'omico Nacional, Instituto de
Astronom\'{\i}a, Universidad Nacional Aut\'onoma de M\'exico, A.P.
877, Ensenada, BC 22860, Mexico.}

\altaffiltext{2}{Department of Astronomy, Beijing Normal University,
Beijing 100875, China.}

\altaffiltext{3}{Instituto de Astrof\'{\i}sica de Canarias,
C/V\'{\i}a L\'actea s/n, E-38205, La Laguna, Tenerife, Spain.}

\shortauthor{Fox Machado et al.}
\shorttitle{Asteroseismology of V650 Tauri}

\listofauthors{L. Fox Machado, et al.}
\indexauthor{Fox Machado, L.}

\abstract{ The preliminary results of a multisite photometric
campaign on the Pleiades Delta Scuti variable V650 Tauri are
reported. The star was observed photometrically for 14 days during
2008 November from three observatories distributed in Longitude
around the Earth. As a result of the preliminary analysis we have
detected in V650 Tauri at least nine oscillation frequencies above a
99\% confidence level.}

\resumen{Presentamos los resultados preliminares de una campa\~na
observacional multisitio sobre  la estrella tipo $\delta$ Scuti V650
Tauri perteneciente al c\'umulo de las Pl\'eyades. La estrella fue
observada durante 14 dias en Noviembre de 2008 desde tres
observatorios distribuidos en longitud alrededor de la Tierra. Como
resultado del an\'alisis preliminar nueve frecuencias de
oscilaci\'on han sido detectadas en V650 Tauri con un nivel de
confidencia superior al 99\%.}

\addkeyword{Stars: $\delta$ Sct}
\addkeyword{Stars: oscillations}

\begin{document}
\maketitle

\section{Introduction}
\label{sec:intro}

Stellar oscillations provide a powerful tool for studying the
interiors of the stars since the mode frequencies depend on the
properties of the star and give strong constraints on stellar models
and hence evolution theories. However, the observations of stellar
pulsations require extensive data sets in order to achieve accurate
frequencies and to avoid the side-lobes in the amplitude spectrum
caused by the daily cycle. Long time series are usually obtained
from the ground by means of multisite observations.

A good scenery to carry out seismic studies are short period
pulsating stars in open clusters. Since the cluster members have
been formed simultaneously in the past they share similar stellar
properties in the present as age, chemical composition and distance.
By means of isochrone fitting it is possible to fix the age and
stellar masses. These constraints imposed by the cluster are very
useful in computing seismic models [e.g. \citet{fox4,fox3,suarez}].
This has motivated, for instance, a number of observational studies
on the Pleiades $\delta$ Scuti stars.  In particular,
  six $\delta$ Scuti variables have been discovered
in the Pleiades cluster until now and most of them have
   been  intensively observed in recent years namely: V647 Tauri
 \citep{liu}, HD 23628 \citep{li1}, V534 Tauri \citep{fox5, li2},  V624
Tauri \citep{fox1, fox2}, HD 23194 \citep{fox1, fox2} and V650 Tauri
\citep{kim}.

The target star V650 Tau (HD 23643, $V=7^{\rm m}.79$, A7)  was
identified as a short-period pulsating variable by \citet{breger1}.
 One-site CCD
photometric observations carried out by \citet{kim} in
November-December 1993, detected four frequencies.
 With the aim at detecting more pulsation frequencies that may be helpful
 in constructing new seismic model for the star \citep{fox6} we have organized
 a multisite campaing on V650 Tauri.

\begin{figure*}[!ht]
  \centering
  \subfigure[]{\includegraphics[width=9cm]{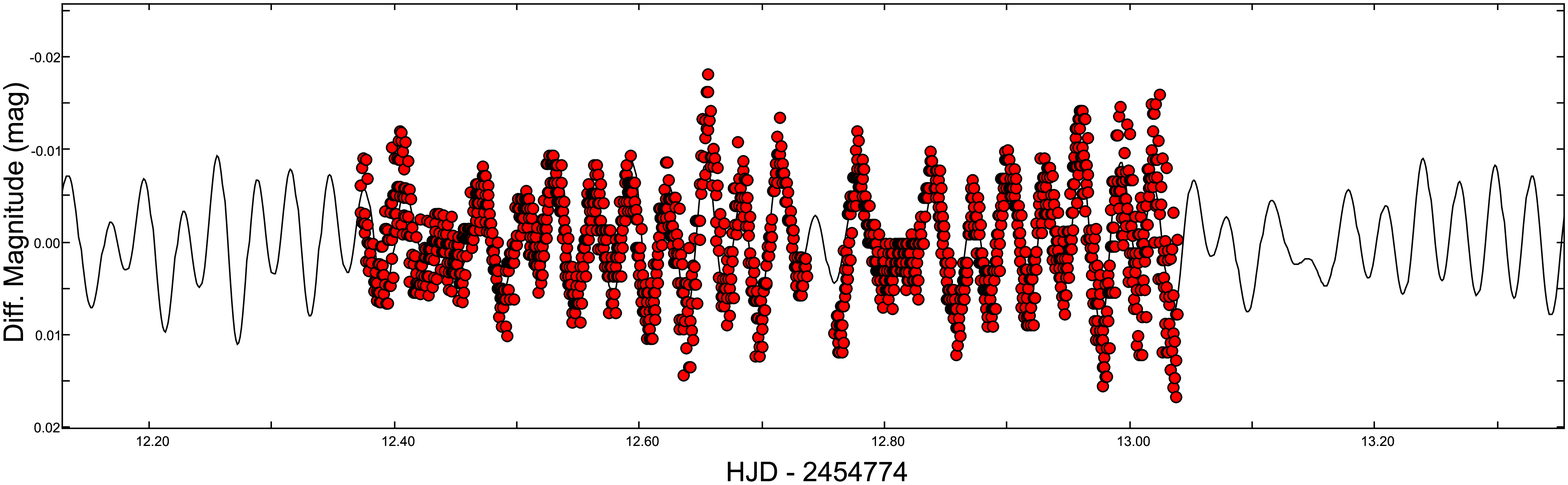}}
  \subfigure[]{\includegraphics[width=7cm]{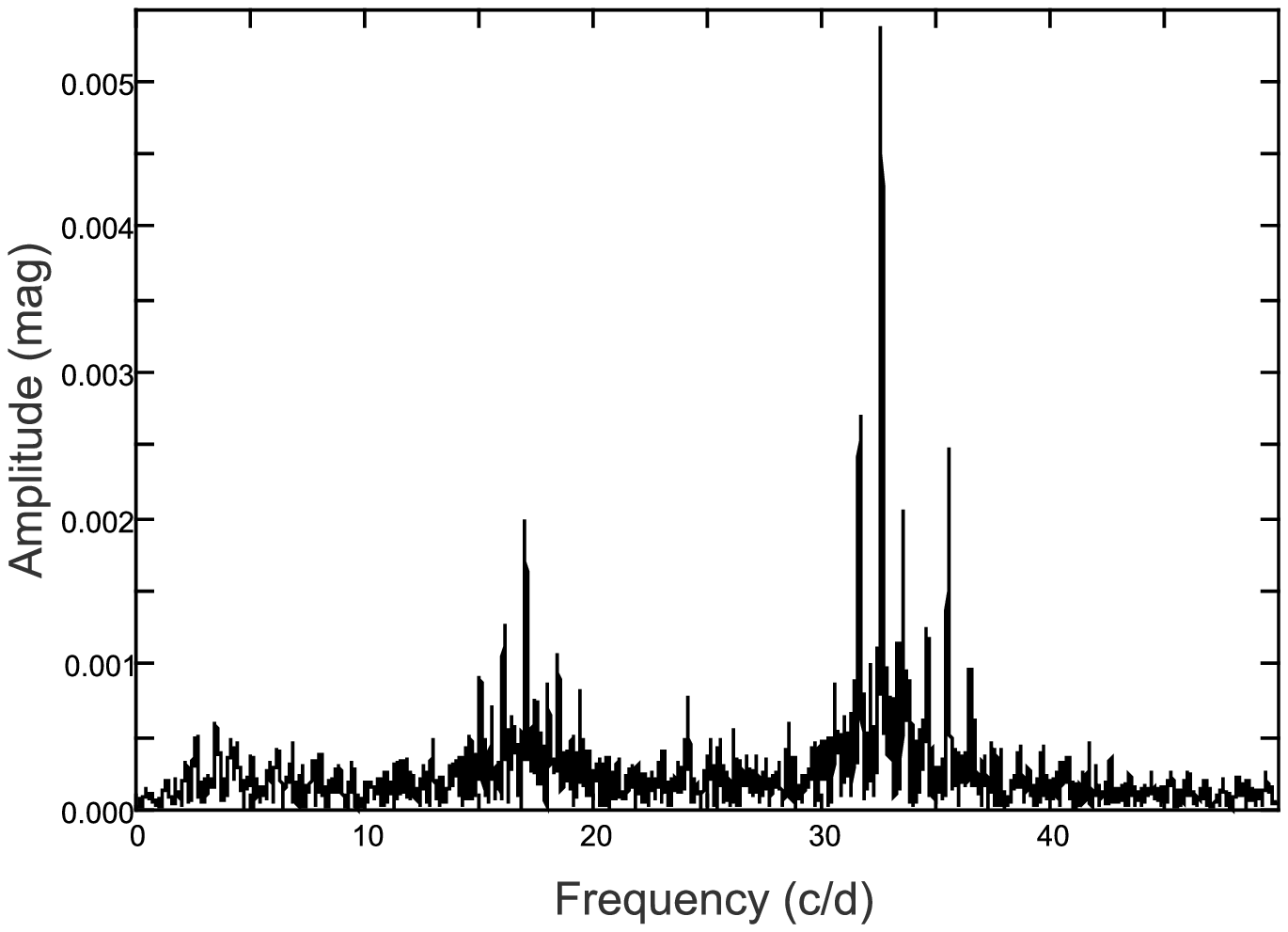}}
  \caption{(a) Example of the differential light curve V650 Tauri - Comparison. The solid line represents a fit to the
  observed data of
  the nine frequency detected in V650 Tauri. (b) Amplitude spectrum derived from the light curve V650 Tau - Comparison. The amplitude is in mag and
 the frequencies in c/d.}
 \label{fig:curva}
\end{figure*}

\begin{table}[!t]\centering
 \caption{Frequency peaks detected in a preliminary analisys of the light curve V650 Tauri - Comp. S/N
is the signal-to-noise ratio in amplitude after the prewhitening
process.} \label{tab:frec}
  \begin{tabular}{lccr}
\hline
Freq.&  A & $\varphi$/($2\pi$) & $S/N$  \\
(c/d)&(mag)&&\\
\hline
32.623471& 0.005075 &0.910606&34.3\\
 35.565042 &0.002194 & 0.777990&17.6 \\
17.023812&0.002064 & 0.305423&8.9\\
18.460618&0.001216 & 0.659008&5.7\\
15.030028 &0.000832 &0.529173&4.2\\
32.705001 & 0.000761 & 0.605440&5.2\\
3.4312070 &0.000646 & 0.249472&4.0\\
24.177849 &0.000621 & 0.466309&4.6\\
32.862220 & 0.000672 &0.144418&4.6\\
4.1768560 & 0.000598& 0.383873&2.9\\
 \hline
\end{tabular}
\end{table}

\section{Observations and data reduction}
\label{sec:obs}

Three site observations were obtained with a 0.50-m telescope of
Xing Long Station of National Astronomical Observatory of China, the
 0.84-m telescope of
Observatorio San Pedro M\'artir in Mexico and the 0.80-m telescope
IAC80 of Teide Observatory in Spain. For all these observations CCD
cameras  were used. The photometric data at Teide observatory and
Xing Long Station were acquired through a Johnson $V$ filter. In San
Pedro M\'artir a Str\"omgren $y$ filter was used instead. More than
16000 frames on 21 nights were obtained. All data were reduced using
standard IRAF routines. Aperture photometry was implemented to
extract the instrumental magnitudes of the stars. The differential
magnitudes were normalized by subtracting the mean of differential
magnitudes for each night. An example of the differential light
curves is shown in Figure~\ref{fig:curva}(a).

\section{Period analysis}

The period analysis has been performed by means of standard Fourier
analysis and least-squares fitting. In particular, the amplitude
spectra of the differential time series were obtained by means of
Period04 package \citep{lenz}, which considers Fourier as well as
multiple least-squares algorithms. This computer package allows to
fit all the frequencies simultaneously in the magnitude domain

The amplitude spectrum  of the differential light curve V650 Tauri -
Comparison  is shown in Figure~\ref{fig:curva}(b).  As can be seen,
V650 Tauri presents high-amplitude peaks distributed between 12 c/d
and 38 c/d.

The frequencies have been extracted by means of standard
prewhitening method. In order to decide which of the detected peaks
in the amplitude spectrum can be regarded as intrinsic to the star
we follow  Breger's criterion given by \citet{breger2},  where it
was shown that the signal-to-noise ratio (in amplitude) should be at
least  4 in order to ensure that the extracted frequency is
significant.

The frequencies, amplitudes and phases  are listed in
Table~\ref{tab:frec}. Nine significant frequencies have been
detected in V650 Tauri in this preliminary analysis.
 A detailed analysis of these observations will be given in a forthcoming
paper.

\medskip
We would like to thank the staff of the San Pedro M\'artir and Teide
observatories and Xing Long station for their assistence in securing
the observations. This work was partially supported by PAPIIT
IN114309.


\begin{thebibliography}
\bibitem[Breger (1972)]{breger1} Breger, M.\ 1972, \apj, 176, 367
\bibitem[Breger et al. (1993)]{breger2} Breger, M. et al.\ 1993, \aap, 271, 482
\bibitem[Fox Machado et al. (2000)]{fox5} Fox Machado, L., et al.\
2000, ASP Conf. Ser., 203, 477
\bibitem[Fox Machado et al. (2001)]{fox4} Fox Machado, L., et al.\
2001, ESA-SP 464, 427
\bibitem[Fox Machado et al. (2002)]{fox1} Fox Machado, L., et al.\
2002, \aap, 382, 556
\bibitem[Fox Machado (2003)]{fox6} Fox Machado, L.,\ 2003, PhD
Thesis, Universidad de La Laguna
\bibitem[Fox Machado et al. (2006)]{fox3} Fox Machado, L., et al.\
2006, \aap, 446, 611
\bibitem[Fox Machado et al. (2008)]{fox2} Fox Machado, L., et al.\
2008, CoAst, 157, 307
\bibitem[Kim \& Lee (1996)]{kim} Kim, S.-L., \& Lee, S.-W.\ 1996, \aap, 310, 831
\bibitem[Lenz \& Breger (2005)]{lenz}Lenz, P., \& Breger, M.\ 2005, CoAst, 146, 53
\bibitem[Li et al. (2002)]{li1} Li, Z.P., et al.\ 2002, \aap, 395,
873
\bibitem[Li et al. (2004)]{li2} Li, Z.P., et al.\ 2004, \aap, 420,
283
\bibitem[Liu et al. (1999)]{liu} Liu, Y.Y., et al.\ 1999, Chinesse
\aap, 23, 349
\bibitem[Su\'arez et al. (2002)]{suarez} Su\'arez, J.C., et al.\
2002, \aap, 390, 523





\end{thebibliography}
\end{document}